\documentstyle[aps]{revtex}
\begin{document}
\draft
\title{
Multifractal Behavior of the Two-Dimensional Ising Model 
at small spatio-temporal scales
\cite{N}}
\author{
W. Sakikawa and O. Narikiyo}
\address{
Department of Physics, 
Kyushu University, 
Fukuoka 810-8560, 
Japan}
\date{
December, 2001}
\maketitle
\begin{abstract}
The distribution of the fractal dimension 
of the two-dimensional Ising model 
at the critical temperature 
measured by the Monte-Carlo simulation is discussed. 
At small spatio-temporal scales 
it exhibits a multifractal behavior 
and is well fitted by a non-Gaussian distribution 
derived from the R\'enyi or Tsallis entropy. 
\end{abstract}
\vskip 18pt
  Recently a large-deviation analysis\cite{FT} 
for the Monte-Carlo data 
of the two-dimensional Ising model at the critical temperature 
has been attempted. 
  Inspired by this attempt we perform a multifractal analysis 
for the same model in this Short Note. 
  Here we do not distinguish the large-deviation and the multifractal 
schemes.\cite{EM} 

  Multifractal behaviors of the Ising model are expected 
for random or chaotic lattices,\cite{OY} 
while the model on a regular lattice is understood 
in terms of the mono-fractal analysis.\cite{IS} 
  Even for the model on a regular lattice, 
we can expect multifractal behaviors when we observe it 
at small spatio-temporal scales. 
  This is a view point of the large-deviation scheme.\cite{FI} 

  In the following we measure the distribution 
of the fractal dimension of the model on the square lattice 
at the critical temperature and show that the data is fitted 
by a non-Gaussian distribution 
derived from the R\'enyi or Tsallis entropy. 
  Such a non-Gaussian behavior originates from 
the multifractality of the spin states 
and the $q$-index appearing in the distribution 
serves as a measure for the multifractality. 
  On the other hand 
the non-Gaussian behavior is observed 
at small spatio-temporal scales 
as expected from the large-deviation scheme 
so that the $q$-index also serves as a measure 
for the degree of non-equilibrium. 

  We measure the fractal dimension of the Ising model 
on the square lattice, 
\begin{equation}
H = -J \sum_{<ij>}^{L \times L} \sigma_i \sigma_j, 
\end{equation}
where we adopt the ordinary notation, 
$<ij>$ means the nearest neighbor bond 
and $L$ is the number of the spins in $x$- or $y$-direction. 
  The fractal dimension,\cite{IS} 
\begin{equation}
D = \ln[M(1)/M(b)]/\ln(b),  
\end{equation}
is determined by the Monte-Carlo data at the critical temperatute, 
$T_{\rm c} = 2J / \ln(1+\sqrt{2}) k_{\rm B}$. 
  Here $b$ is the scaling parameter for the coarse-graining 
and $L \times L$ spins are transformed into $(L/b)\times(L/b)$ block spins. 
  $M(b)$ is the total spin at the level of $b$ of the coarse-graining. 
  The Monte-Carlo up-dating is done with the Metropolis algorithm. 
  The simulation time, $t$, is represented 
by the unit of the usual Monte-Carlo step (MCS). 
  In the following 
the measurement is done during $ \tau < t < \tau + T$. 
  The simulation is performed on our personal computer 
with Pentium-4 CPU. 

  Some typical data for the unnormalized distribution 
of the fractal dimension are shown in Fig.\ 1 
where $D$ is measured at each Monte-Carlo steps 
and the total number of the measurement is $T$. 
  The simulation data are well fitted by the distribution function, 
\begin{equation}
P(D) = P_0 \cdot [1 + (q-1)(D - D_0)^2/2\sigma^2]^{1/(1-q)},  
\end{equation}
which is derived by the maximum-entropy principle 
using the R\'enyi or Tsallis entropy\cite{Tsallis,AA,SB} 
so that the distribution exhibits a multifractality of the system. 
  The Tsallis entropy is non-extensive, 
while the R\'enyi entropy is extensive. 
  The functional form of the distribution is independent of 
the choice of the entropy, the R\'enyi or Tsallis,\cite{AA} 
so that our result is independent of the extensivity of the entropy. 
  Here $q$ represents the multifractality of the system investigated 
and the mean value, $D_0 \sim 1.875$, is the one 
expected from the mono-fractal analysis.\cite{IS} 
  The limiting case with $q \rightarrow 1$ 
corresponds to the Gaussian distribution 
which is parabolic in the $D$-$\ln P$ plane. 
  The $q$-index for the solid curve is determined 
by the least-square fitting. 
  In our scheme 
$\sigma$ is also determined by the data fitting. 
  The peak value of the unnormalized distribution is $P_0$. 

  In Fig.\ 2 
the system-size dependence of the $q$-index is shown. 

  In Fig.\ 3 
the simulation-time dependence of the $q$-index is shown. 

  From the data in Figs.\ 2 and 3 
we can read a tendency that $q \rightarrow 1$ 
at large spatio-temporal scales. 
  The value, $q=1$, is expected 
for the equilibrium state in the thermodynamic limit. 
  Our data at small spatio-temporal scales 
reflect a deviation from this limit 
and exhibit non-equilibrium nature. 
  A similar analysis has already been done 
for the experimantal data for turbulene.\cite{Tsallis} 

  In conclusion 
we numerically observed a multifractal behavior 
of the two-dimensional Ising model 
at small spatio-temporal scales. 
  The $q$-index serves as a measure not only for multifractality 
but also non-equilibrium in this case. 

\vskip 100pt

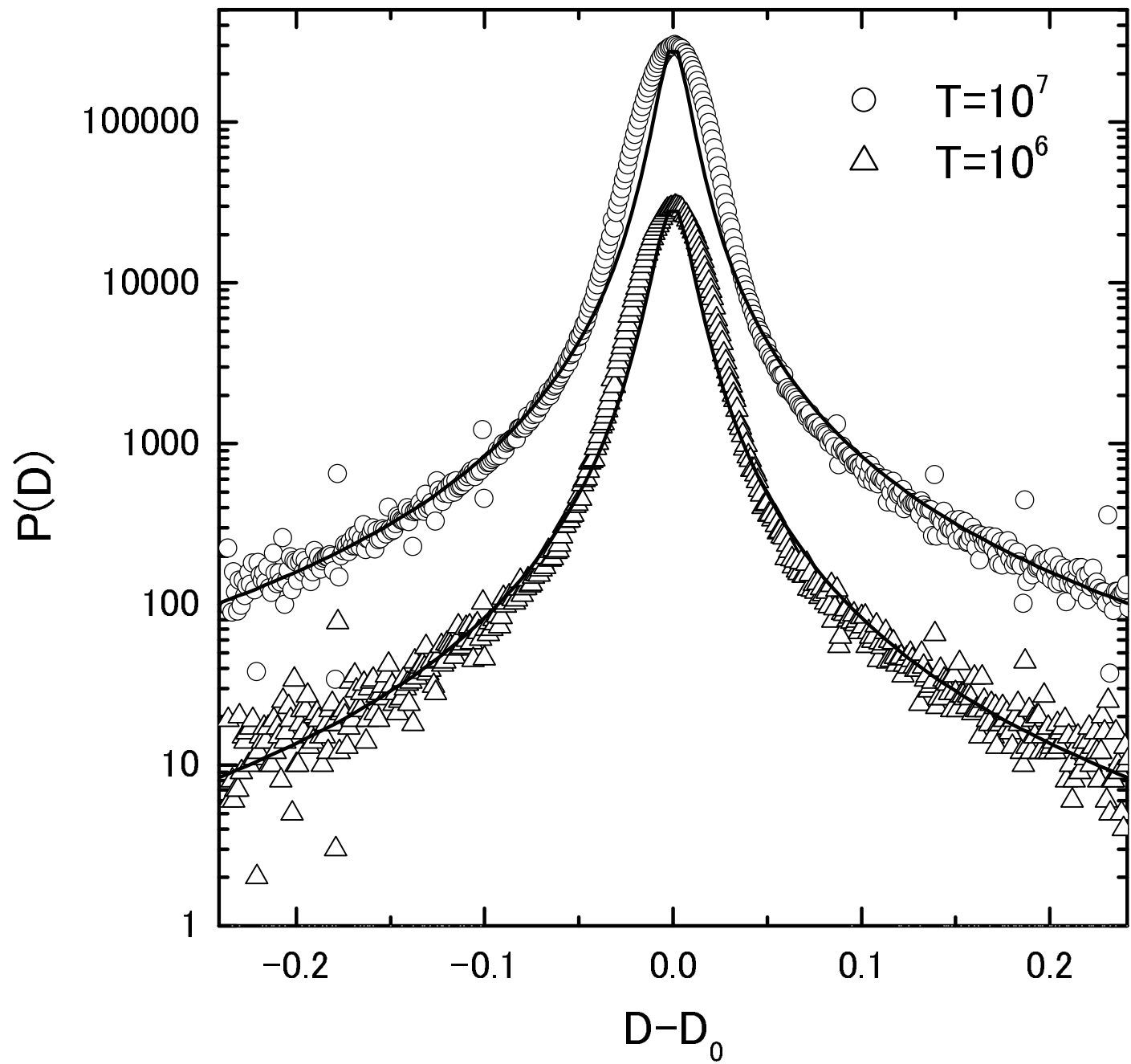
\begin{figure}
\caption{
The distripution of the fractal dimension 
for a randomly-produced initial state 
with $L=125$, $b=5$ and $\tau = 10^5$. }
\label{fig:1}
\end{figure}

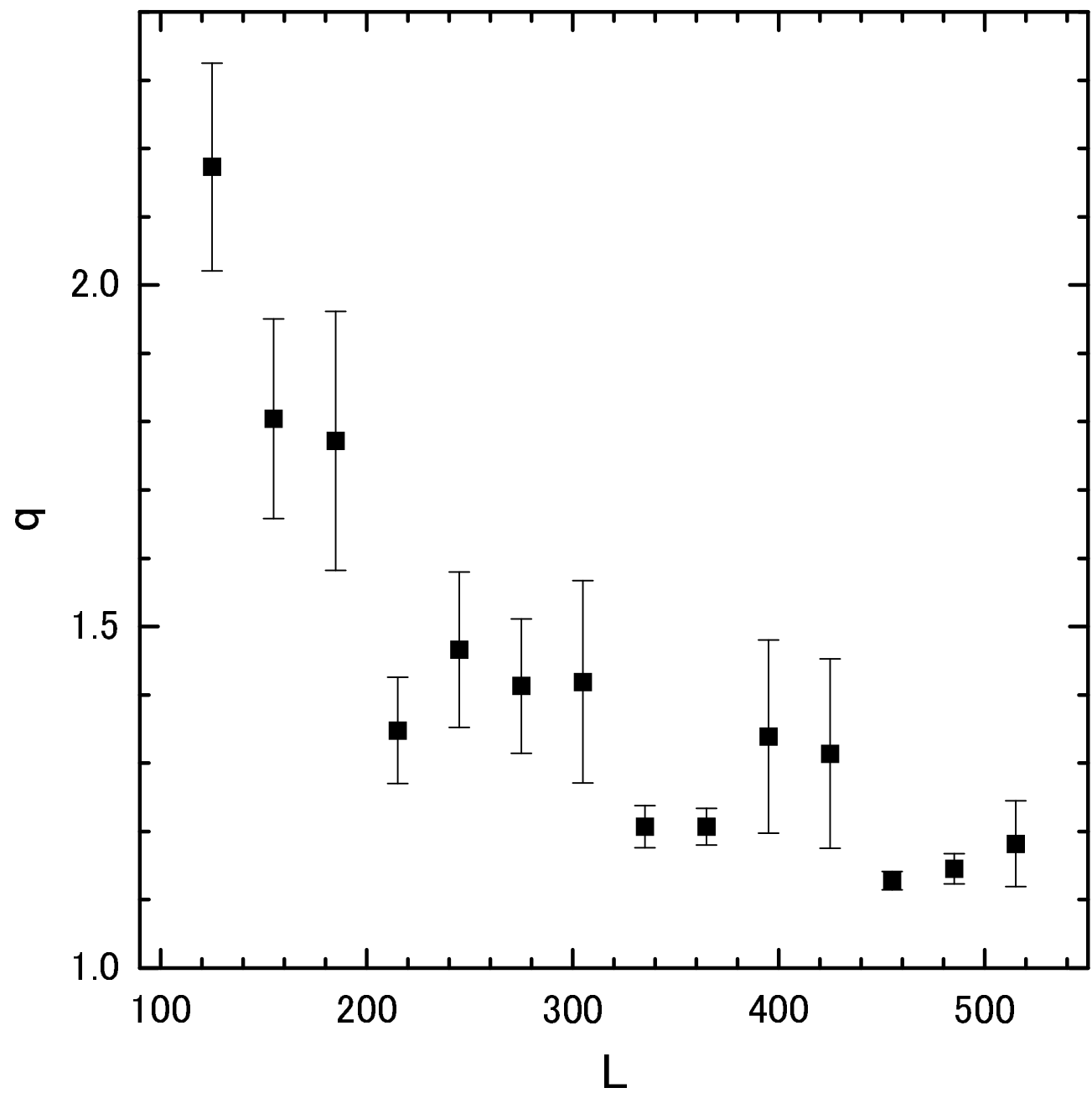
\begin{figure}
\caption{
The system-size dependence of the $q$-index 
averaged over 10 randomly-produced initial states 
with $b=5$, $\tau = 10^5$ and $T = 10^4$. }
\label{fig:2}
\end{figure}

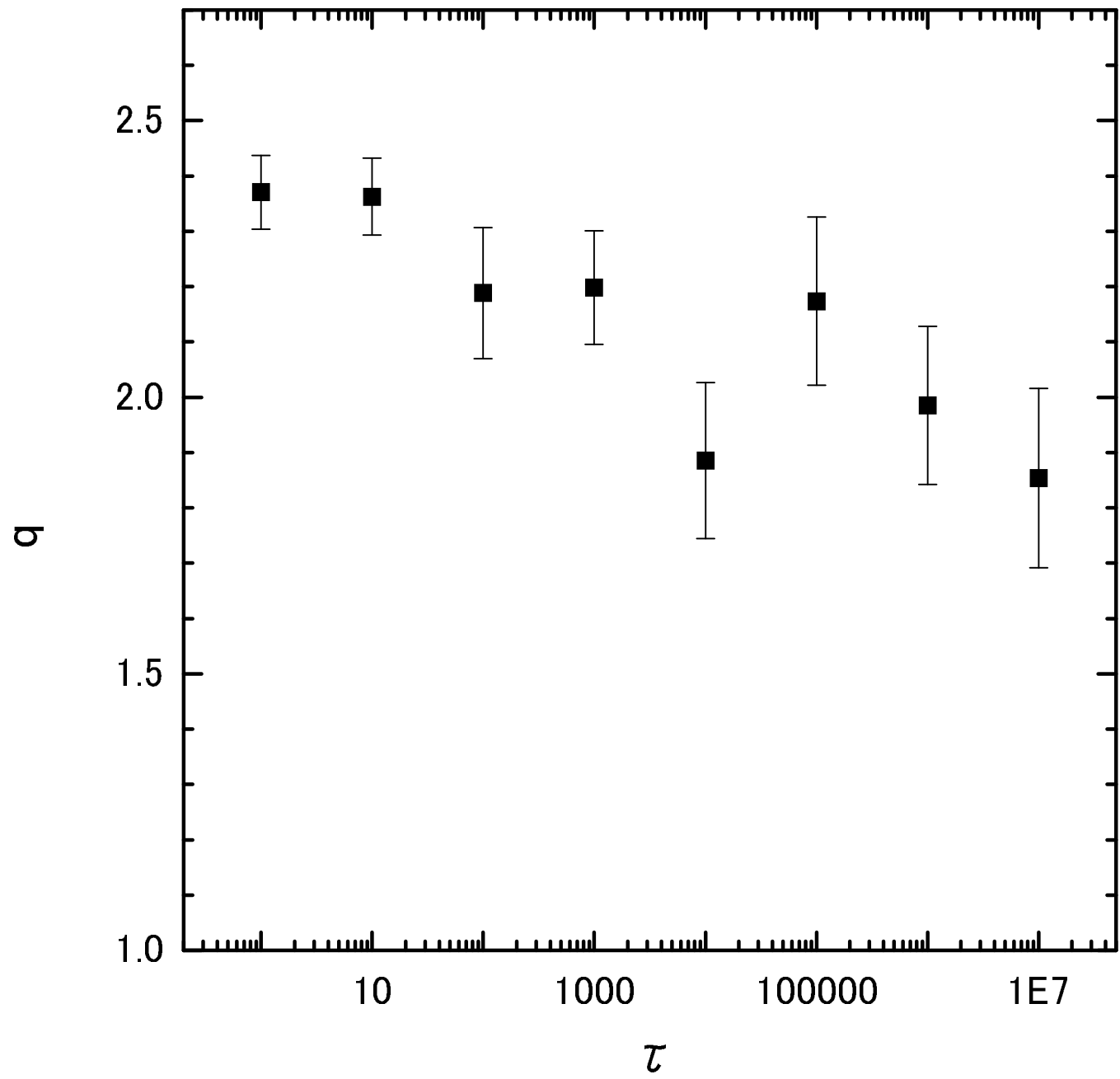
\begin{figure}
\caption{
The simulation-time dependence of the $q$-index 
averaged over 10 randomly-produced initial states 
with $L=125$, $b=5$ and $T = 10^5$. }
\label{fig:3}
\end{figure}


\begin{references}

\bibitem{N}
This paper has been submitted to the Short Note section 
of J. Phys. Soc. Jpn.

\bibitem{FT}
H.\ Fujisaka and H.\ Tutu: Unpublished work 
where the coarse-grained order-parameter measured 
by the Monte-Carlo simulation is analyzed 
in terms of the entropy function of the large-deviation theory. 

\bibitem{EM}
C.\ J.\ G.\ Evertsz and B.\ B.\ Mandelbrot: 
{\it Chaos and Fractals} H-O.\ Peitgen, H.\ J\"urgens and D.\ Saupe 
(Springer-Verlag, New York, 1992) p.921. 

\bibitem{OY}
T.\ Olson and A.\ P.\ Young: 
Phys. Rev. B {\bf 60} (1999) 3428 and references therein. 

\bibitem{IS}
N.\ Ito and M.\ Suzuki: 
Prog. Theor. Phys. {\bf 77} (1987) 1391. 

\bibitem{FI}
H.\ Fujisaka and M.\ Inoue: 
Phys. Rev. A {\bf 39} (1989) 1376. 

\bibitem{Tsallis}
C.\ Tsallis: 
{\it Nonextensive Statistical Mechanics and Its Applications} 
S.\ Abe and Y.\ Okamoto (Springer-Verlag, Berlin, 2001) p.3. 

\bibitem{AA}
T.\ Arimitsu and N.\ Arimitsu: 
Physica A {\bf 295} (2001) 177. 

\bibitem{SB}
B.\ K.\ Shivamoggi and C.\ Beck: 
J. Phys. A: Math. Gen. {\bf 34} (2001) 4003. 

\end{references}
\end{document}